\documentclass[11pt,a4paper]{article}

\usepackage[dvips]{graphicx}
\graphicspath{{eps/}}

\usepackage[cp1251]{inputenc}
\usepackage[T2A]{fontenc}
\usepackage[russian]{babel}

\usepackage{animate}

\usepackage{amssymb}
\usepackage{amsxtra}
\usepackage{amsthm}
\usepackage{color}
\usepackage{hyperref}
\numberwithin{equation}{section}
\theoremstyle{plain}
\newtheorem{theorem}{Теорема}

\theoremstyle{remark}

\def\sgrad{\mathrm{sgrad}\,}
\def\trace{\mathrm{trace}\,}
\newcommand{\rank}{\mathop{\rm rank}\nolimits}

\begin{document}

\title{Bifurcation diagram of one perturbed vortex dynamics problem}

\author{P.\,E.~Ryabov}

\date{}

\maketitle

\begin{abstract}
This paper deals with the one perturbed vortex dynamics problem which describes the system of two point vortices in a Bose-Einstein condensate enclosed in a cylindrical trap. This system is a completely integrable Hamiltonian system with two degrees of freedom.   Bifurcation diagram of momentum mapping is analytically investigated.  To find the bifurcation diagram we use the method of critical subsystems developed by M.P. Kharlamov for the study of the phase topology of integrable problems of rigid body dynamics. As an application, we present an analysis of the stability of critical trajectories (i.e., the nondegenerate singularities of rank 1 of the momen\-tum mapping) by defining a trajec\-tory type (elliptic/hyperbolic) for each curve from the bifurcation set.

\vspace{3mm}


\end{abstract}
{

\parindent=0mm

УДК 532.5.031, 517.938.5


------------------------------------------------------------

Received on 23 November 2018.

------------------------------------------------------------

The work was supported by RFBR
grants 16-01-00170 and  17-01-00846.

------------------------------------------------------------

Ryabov Pavel Evgen`evich

PERyabov@fa.ru

Financial University under the Government of the Russian Federation\\
Leningradsky prosp. 49, Moscow, 125993 Russia

Institute of Machines Science, Russian Academy of Sciences\\
Maly Kharitonyevsky per. 4, Moscow, 101990 Russia

Udmurt State University\\
ul. Universitetskaya 1, Izhevsk, 426034 Russia

---------------------------------------------------------------
}

\tableofcontents

\section{Введение}

 В настоящей работе рассматривается система двух вихревых нитей в бо\-зе-эйн\-штей\-нов\-ском кон\-ден\-са\-те, заключенном в цилиндрической ловушке (см. \cite{kevrek2011}). Динамика системы двух точечных вихрей описывается системой обыкновенных дифференциальных уравнений относительно координат вихревых нитей, которая может быть представлена в гамильтоновой форме
\begin{equation}
\label{x1}
\boldsymbol{\dot{\zeta}}=\{\boldsymbol\zeta, H\}
\end{equation}
c функцией Гамильтона
\begin{equation}
\label{x2}
H=\ln[1-(x_1^2+y_1^2)]+a^2\ln[1-(x_2^2+y_2^2)]-a b\ln[(x_2-x_1)^2+(y_2-y_1)^2].
\end{equation}
Здесь через $(x_k,y_k)$ обозначены декартовы координаты $k$-ого вихря ($k=1,2$), фазовый вектор $\boldsymbol\zeta$ имеет координаты $\{x_1,y_1,x_2,y_2\}$, параметр $a$ обозначает отношение интенсивностей $\frac{\Gamma_2}{\Gamma_1}$. Физический параметр $b$ характеризует меру вихревого взаимодействия интенсивностей и пространственной неоднородности в гармонической  ловушке \cite{kevrek2011}, \cite{kevrek2013}. В работах \cite{SokRyabRCD2017}, \cite{sokryab2018} такой параметр принимался равным единице, тем не менее в ряде физических работ для параметра $b$ в случае равных интенсивнойстей ($a=1$) на основе экспериментальных данных принимались следующие значения $b=2$, $b=1,35$, $b=0,1$  и др.

Фазовое пространство $\cal P$ задается в виде прямого произведения двух открытых кругов радиуса $1$, с выколотом множеством столкновениий вихрей
\begin{equation*}
\label{x3}
{\cal P}=\{(x_1,y_1,x_2,y_2)\,:\, x_1^2+y_1^2<1, x_2^2+y_2^2<1\}\smallsetminus \{x_1=x_2, y_1=y_2\}.
\end{equation*}
Пуассонова  структура на фазовом пространстве $\cal P$ задается в стандартном виде
\begin{equation*}
\label{x4}
\{x_i,y_j\}=-\frac{1}{\Gamma_i}\delta_{ij},
\end{equation*}
где $\delta_{ij}$ -- символ Кронекера.

Система \eqref{x1} допускает один дополнительный первый интеграл движения -- \textit{момент завихренности}
\begin{equation*}
\label{x5}
F=x_1^2+y_1^2+a(x_2^2+y_2^2).
\end{equation*}

Функция $F$ вместе с гамильтонианом $H$ образуют на $\cal P$ полный инволютивный набор интегралов системы $\eqref{x1}$.
Согласно теореме Арнольда–Лиувилля можно утверждать, что компактная связная компонента интегрального многообразия ${\cal M} = \{H = h, F = f\}$ диффеоморфна двумерному тору. Определим \textit{интегральное отображение} ${\cal F}\,:\, {\cal P}\to {\mathbb R}^2$, полагая $(f,h)={\cal F}(\boldsymbol\zeta)=(F(\boldsymbol\zeta), H(\boldsymbol\zeta))$. Отображение $\cal F$ принято также называть \textit{отображением момента}.
Обозначим через $\cal C$ совокупность всех критических точек отображений момента, то есть
точек, в которых $\rank d{\cal F}(x) < 2$. Множество критических значений $\Sigma = {\cal F}({\cal C}\cap{\cal P})$ называется \textit{бифуркационной диаграммой}.
В работах \cite{SokRyabRCD2017,sokryab2018} для случая, когда параметр взаимодействия $b$ полагался равным единице,  исследована фазовая топология динамики двух вихревых нитей и их динамические эффекты.
Основной целью данной статьи является явное определение бифуркационной диаграммы для возмущенной задачи, т.е. когда параметр взаимодествия $b$ принимает любые положительные значения. Для физических приложений особый интерес представляет случай, когда интенсивности положительны и равны (т.е. $a=1$), а параметр $b$ отличен от единицы. Обнаружены новые свойства бифуркационной диаграммы, которые ранее не встречались в задачах вихревой динамики \cite{sokryab2018, bormam2005}.

 Для нахождения бифуркационной диаграммы мы используем метод критических подсистем, развитый М.~П.~Харламовым  при исследовании фазовой топологии интегрируемых задач динамики твердого тела. При анализе устойчивости невырожденных (в смысле теории особенностей) траекторий бифуркационная диаграмма  позволяет быстрым и наглядным образом определять устойчивость в тех случаях, когда использование общих стандартных методов является довольно затруднительным. В качестве приложения приводится  анализ устойчивости критических траекторий (т.е. невырожденных особенностей ранга 1 отображения момента)  путем определения типа  траектории (эллиптический/ги\-пер\-бо\-ли\-че\-ский) для каждой кривой из бифуркационного множества.

\section{Критическая подсистема и бифуркационная\\ диаграмма отображения момента}
Определим следующие полиномиальные выражения $F_k$ от фазовых переменных
\begin{equation*}
\label{y1}
\begin{array}{l}
F_1=x_1y_2-y_1x_2,\\[3mm]
F_2=a(x_2^2+y_2^2)^2x_1^4-(x_2^2+y_2^2)[(x_2^2+y_2^2)(a-b)+b]x_2x_1^3+\\[3mm]
+(x_2^2+y_2^2)[(ab-1)(x_2^2+y_2^2-1)-a]x_1^2x_2^2+\\[3mm]
[(x_2^2+y_2^2)(x_2^2+y_2^2+a-b-1)+b]x_2^3x_1-ab(x_2^2+y_2^2-1)x_2^4
\end{array}
\end{equation*}
и обозначим через ${\cal N}$ замыкание множества решений системы
\begin{equation}
\label{y2}
F_1 = 0,\quad F_2 = 0.
\end{equation}

Тогда справедлива теорема.

\begin{theorem}
\label{t1}
Множество $\cal C$ критических точек  отображения момента $\cal F$ совпадает с множеством решений системы \eqref{y2}. Множество ${\cal N}$  является двумерным инвариантным подмногообразием системы \eqref{x1} с гамильтонианом \eqref{x2}.
\end{theorem}

\begin{proof}
Для доказательства первого утверждения теоремы\linebreak необходимо найти
точки фазового пространства, в которых ранг отображения момента не максимален. С помощью
прямого вычисления можно убедиться, что матрица Якоби отображения момента имеет
нулевые миноры второго порядка в точках $\zeta\in{\cal P}$, координаты которых удовлетворяют уравнениям системы \eqref{y2},
откуда ${\cal C} = {\cal N}$. В инвариантности соотношений \eqref{y2} можно убедиться с помощью следующей цепочки верных равенств
\begin{equation*}
\dot F_1 = \{F_1,H\}_{F_1=0} = \sigma_1F_2,\quad \dot F_2 = \{F_2,H\}_{F_1=0} = \sigma_2F_2,
\end{equation*}
где $\sigma_k$ некоторые полиномиальные функции от фазовых переменных.
\end{proof}

Для определения бифуркационной диаграммы $\Sigma$ удобно перейти к полярным координатам с помощью соотношений
\begin{equation*}
\label{y4}
\begin{array}{l}
x_1 = r_1\cos\theta_1,\quad y_1 = r_1\sin\theta_1,\\[3mm]
x_2 = r_2\cos\theta_2,\quad y_2 = r_2\sin\theta_2.
\end{array}
\end{equation*}

Первое из уравнений системы \eqref{y2} принимает вид $\sin(\theta_1-\theta_2)=0$, т.е. $\theta_1-\theta_2=0$ и $\theta_1-\theta_2=\pi$. Первая возможность, в отличие от динамики двух вихрей интенсивностей разных знаков \cite{SokRyabRCD2017}, не реализуется ни при каких положительных значениях параметров отношения интенсивностей $a$ и $b$. Для второй возможности, т.е. когда $\theta_1=\theta_2+\pi$, критическое множество $\cal C$ можно представить в виде критической подсистемы ${\cal N}$ в следующем виде
\begin{equation*}
\label{y5}
{\cal N}:\left\{\begin{array}{l}
\theta_1=\theta_2+\pi,\\[3mm]
\gamma:\quad ar_2r_1^4+[(a-b)r_2^2+b]r_1^3+[(ab-1)r_2^3+(1-ab-a)r_2]r_1^2-\\[3mm]
-[r_2^4-(1-a+b)r_2^2+b]r_1+ab(1-r_2^2)r_2=0.
\end{array}\right.
\end{equation*}

Несмотря на то, что род алгебраической кривой $\gamma$ равен четырем, все таки удалось найти приемлемую параметризацию неявной кривой $\gamma$:
\begin{equation*}
\label{x2_3}
\begin{array}{l}
\displaystyle{r_1=\frac{1}{\sqrt{2}}\sqrt{\frac{abt^3+(a-b-1)t^2+(a-1+ab)t-b\pm\sqrt{\cal D}(t+1)}{t[-t^3+(ab-1)t^2+(a-b)t+a]}}},\\[5mm]
r_2=t\cdot r_1, \text{\quad где\quad}{\cal D}=[abt^2+(a+1)(1-b)t+b]^2-4at^2.
\end{array}
\end{equation*}
Соответствующая бифуркационная диаграмма $\Sigma$ задается в виде кривой на плоскости ${\mathbb R}^2(f,h)$:
\begin{equation}
\label{x2_4}
\Sigma_{a,b}:\left\{
\begin{array}{l}
f=(1+at^2)r_1^2,\\[3mm]
h=\ln(1-r_1^2)+a^2\ln(1-t^2r_1^2)-ab\ln[(1+t)^2r_1^2],\\[3mm]
\displaystyle{r_1^2=\frac{abt^3+(a-b-1)t^2+(a-1+ab)t-b\pm\sqrt{\cal D}(t+1)}{2t[-t^3+(ab-1)t^2+(a-b)t+a]}.}
\end{array}\right.
\end{equation}

Особый интерес представляет случай равных положительных интенсивностей, т.е. когда параметр отношения интенсивностей $a$ равен единице. В этом случае бифуркационная диаграмма имеет простой вид и состоит из двух кривых $\Sigma_{1,b}=\gamma_1\cup\gamma_2$, где
\begin{equation}\label{x2_5}
\begin{array}{l}
\displaystyle{\gamma_1: h=2\ln\left(1-\frac{f}{2}\right)-b\ln(2f),\quad 0<f<2;}\\
\gamma_2: \left\{
\begin{array}{l}
\displaystyle{h=\ln\left[\frac{s^2(s-1)}{b+s-1}\right]-b\ln\left[\frac{bs^2}{b+s-1}\right],}\\[3mm]
\displaystyle{f=\frac{bs^2-2(s-1)(b+s-1)}{b+s-1},}
\end{array}\right.\qquad s\in \left(1;\frac{2(1+\sqrt{b})}{2+\sqrt{b}}\right].
\end{array}
\end{equation}
Отметим, что при значениях физического параметра $b>3$, кривая $\gamma_2$ имеет точку возврата при $s_{cusp}=\frac{\left[2-b+\sqrt{b(b-2)}\right](b-1)}{b-2}$, которая совпадает с точкой касания для $b=3$ при $s_{touch}=\frac{2(1+\sqrt{b})}{2+\sqrt{b}}$.

Параметризованная кривая \eqref{x2_4} также имеет точки возврата, которые удовлетворяют уравнению $P(t)=0$. Здесь $P(t)$ -- многочлен десятой степени, коэффициенты  которого зависят от физических параметров $a$ и $b$. Более того, дискриминант этого многочлена описывает ситуацию, когда точки возврата ``сливаются'' в одну и одна из ветвей становится гладкой. Далее, вновь происходит рождение ``точек возврата''.  На рис.~1 и 2 представлены анимации бифуркационных диаграмм, которые соответствуют значениям физического параметра $a=4$ и $a=1$ соответственно, при этом была использована компьютерная система аналитических вычислений.

\begin{figure}[!ht]
\centering
\begin{center}
\animategraphics[width=10cm,height=9cm, controls,buttonsize=1em,buttonfg=0.5]{3}{pic}{001}{12}
\end{center}

\begin{center}
\animategraphics[width=10cm,height=9cm,controls,buttonsize=1em,buttonfg=0.5]{3}{pic_}{001}{60}
\end{center}
\caption{Анимация бифуркационной диаграммы $\Sigma$ для $a=4$.}
\label{fig2}
\end{figure}

\begin{figure}[!ht]
\centering
\begin{center}
\animategraphics[width=10cm,height=9cm,controls,buttonsize=1em,buttonfg=0.5]{3}{pic_a1_}{001}{60}
\end{center}

\begin{center}
\animategraphics[width=10cm,height=9cm,controls,buttonsize=1em,buttonfg=0.5]{3}{cadr_a1_}{001}{60}
\end{center}
\caption{Анимация бифуркационной диаграммы $\Sigma$ для $a=1$.}
\label{fig3}
\end{figure}

\section{Приложение}
В качестве приложения исследуем характер устойчивости критических траекторий, которые лежат в прообразе бифуркационных кривых \eqref{x2_4} и \eqref{x2_5}. При этом достаточно определить тип (эллип\-ти\-че\-ский/ги\-пер\-бо\-ли\-че\-ский) в какой-нибудь одной из точек $(f, h)$ гладкой ветви кривой $\Sigma$ \cite{BolBorMam1}.

Тип критической точки $x_0$ ранга 1 в интегрируемой системе с двумя  степенями свободы вычисляется следующим
образом. Необходимо указать первый интеграл $F$, такой, что $dF(x_0) = 0$ и
$dF\ne 0$ в окрестности этой точки. Точка $x_0$ оказывается
неподвижной для гамильтонова поля $\sgrad F$ и можно вычислить линеаризацию этого поля в данной точке -- оператор $A_F$ в точке $x_0$. Этот оператор будет иметь два нулевых собственных числа, оставшийся сомножитель характеристического многочлена имеет вид $\mu^2 - C_F$, где $C_F=\frac{1}{2}\trace(A_F^2)$. При $C_F < 0$ получим точку типа ``центр'' (соответствующее периодическое решение имеет эллиптический тип, является устойчивым периодическим решением в фазовом пространстве, пределом концентрического семейства двумерных регулярных торов), а при $C_F > 0$ получим точку типа ``седло'' (соответствующее периодическое решение имеет гиперболический тип, существуют движения, асимптотические к этому решению, лежащие на двумерных сепаратрисных поверхностях). Здесь мы предъявим явное выражения для $C_F$ лишь для бифуркационных кривых $\gamma_1$ и $\gamma_2$:
\begin{equation*}\label{x2_6}
\begin{array}{l}
\gamma_1: C_F=(4-b)f^2+4bf-4b,\quad 0<f<2;\\
\gamma_2: C_F=(b-2)s^2+2(b-1)(b-2)s-2(b-1)^2, \quad s\in \left(1;\frac{2(1+\sqrt{b})}{2+\sqrt{b}}\right].
\end{array}
\end{equation*}

На рис.~3 представлен увеличенный фрагмент бифуркационной диаграммы, когда параметр отношения интенсивностей $a$ равен единице, а параметр $b>3$. Знаки $+$ и $-$ соответствуют эллиптическим (устойчивым) и гиперболическим периодическим решениям в фазовом пространстве. Как и следовало ожидать, смена типа происходит в точке касания $B$ и точке возврата $A$ бифуркационной диаграммы $\Sigma$.

\begin{figure}[!ht]
\centering
\includegraphics[width=1\textwidth]{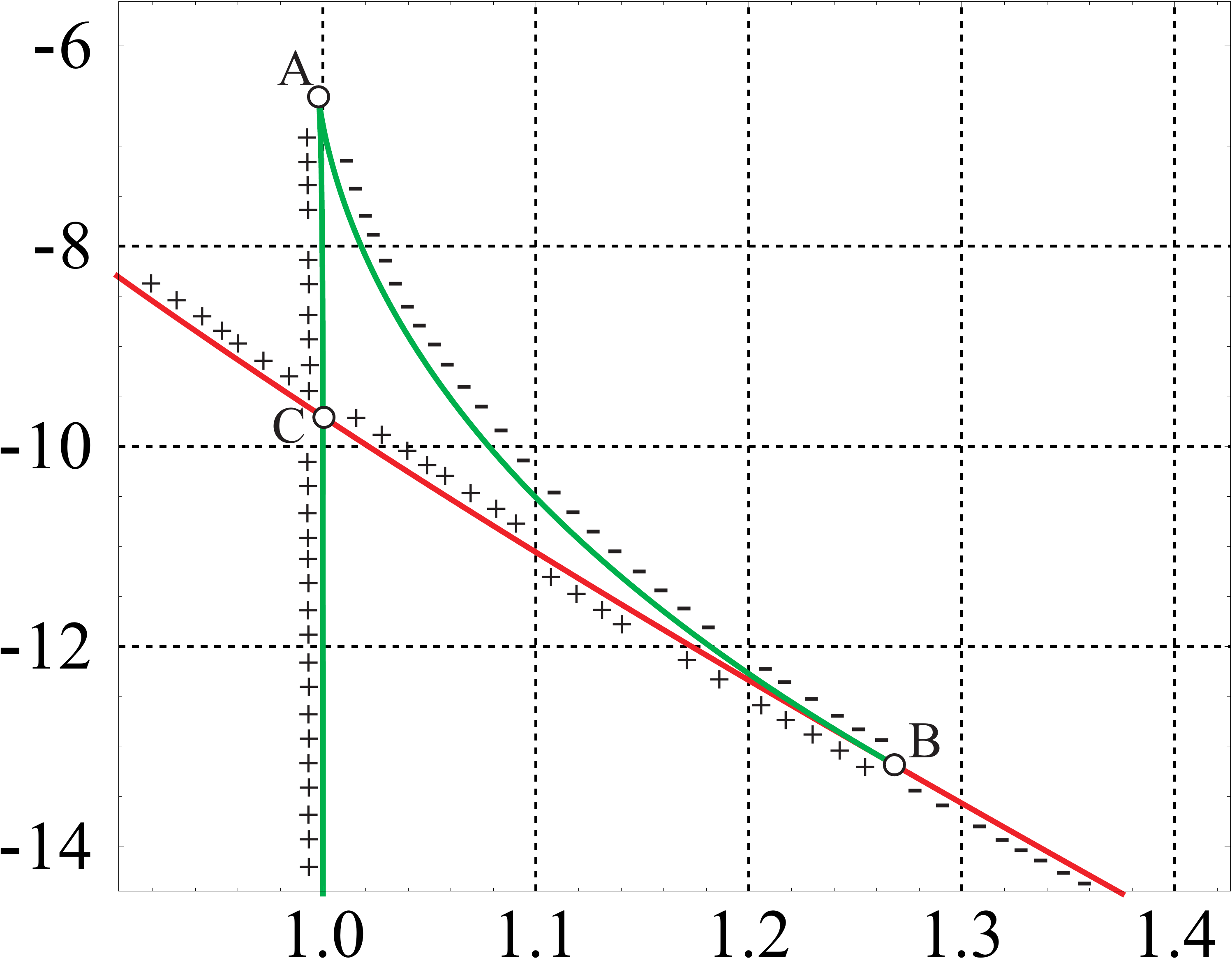}
\caption{Увеличенный фрагмент бифуркационной диаграммы $\Sigma$ для $a=1$ и $b>3$.}
\label{fig4}
\end{figure}

Интерес представляет задача исследования бифуркаций лиувиллевых торов, особенно на участке $(A; B)$ бифуркационной кривой $\gamma_2$. Результаты такого исследования будут опубликованы в отдельной работе.

\section{Благодарности}
Автор выражает благодарность А.\,В.\,Бо\-ри\-со\-ву за  плодотворные обсуждения и ценные советы, касающиеся содержания работы.

\end{document}